\documentclass{aa}
\usepackage[utf8]{inputenc}

\usepackage[varg]{txfonts}
\usepackage{hyperref}
\usepackage{graphicx}
\usepackage{amsmath}
\usepackage{amssymb}
\usepackage{color}

\usepackage{natbib}
\bibpunct{(}{)}{;}{a}{}{,}

\begin{document}

\newcommand{\annot}[1]{({\bf \color{red} #1})}

\title{Ultra Luminous X-ray sources - new distance indicators?}

\author{A.~R\'o\.za\'nska\inst{1}
    \and 
       K.~Bresler\inst{2}
       \and 
       B.~Be\l dycki\inst{1}
       \and 
       J.~Madej\inst{3}
       \and
       T.P.~Adhikari\inst{1}
 }
   \offprints{A.~R\'o\.za\'nska}

\institute{N.~Copernicus Astronomical Center, Bartycka 18, 00-716 Warszawa, Poland \\
          \email{agata@camk.edu.pl}
           \and
           Warsaw University of Technology, Plac Politechniki 1, 00-661 Warszawa,  Poland 
           \and 
        Astronomical Observatory, University of Warsaw, Al. Ujazdowskie 4, 00-478, Warszawa, Poland   
           }

\date{Received September 15, 1996; accepted March 16, 1997}

\abstract
{}
{In this paper we fit the {\it NuSTAR} and {\it XMM-Newton} data of three sources: NGC7793~P13, NGC5907~ULX1, and Circinus~ULX5.}
{Our single model contains emission form non-spherical system: 
 neutron star plus accretion disk directed towards observer.}
{ We obtained the very good fit with the reduced $\chi^2$ per degree of freedom equal 1.08 for P13, 
1.01 for ULX1, and 1.14 for ULX5.
The normalization of our model constrains  the distance to the source. 
The resulting distances are $D=3.41^{+0.11}_{-0.10}$, 
$6.55_{-0.81}^{+0.69}$ and $2.60^{+0.05}_{-0.03}$~Mpc for P13, ULX1 and ULX5 respectively.
The distances to P13 and ULX5 are  in perfect 
agreement with previous distance measurements to their host galaxies.}
{Our results confirm that P13, ULX1 and ULX5 may contain central hot neutron star. When the outgoing emission is computed by integration over the emitting surface and successfully fitted to the data, then the resulting model normalization is the direct distance indicator.}


\maketitle


\section{Introduction}

Recent observations suggest that some Ultra Luminous X-ray (ULX) sources contain
neutron star in their centers, since a typical coherent pulsation from magnetized pulsar 
was detected  in case of: M82X-2 \citep{bachetti14}, NGC5907 ULX1 \citep{israel2017}, and NGC7793 P13 \citep{furst16,israel17}. 
The fact that mass of the central object may equal 1.4 solar masses  implies that 
the accretion disk has to be highly super Eddington. Such a consequence 
directly results from the large values of the observed X-ray luminosities from those 
systems, which always exceed the theoretical maximum for spherical infall 
onto a stellar-mass black hole \citep{roberts07,liu13}. The broad-band spectra of those sources obtained simultaneously 
by {\it XMM-Newton} and {\it NuSTAR} X-ray satellites, in most cases are inadequately
describe by only one multi-temperature disk component. The second spectral component as thermal Comptonization or black body from the neutron star surface is 
needed to fully explain the spectral shape \citep{walton13,WaltonNGC7793_2017}.

Theoretical explanation for the broad spectral shape was given by \citet{rozanska17}
in the case of low mass X-ray binaries (LMXB), where we presented spectral 
shape from non-spherical system which contains neutron star with an accretion disk. 
Our single model component contains emission from both regions with the effects of mutual occultation which gives the proper model normalization. 
Final spectrum depends on the viewing angle in the whole energy range and for the assumed emitting surface it is proportional to the inverse square distance.

In this paper, we reduce archival broad-band data of 
three sources: 
NGC7793~P13, NGC5907~ULX1, and Circinus~ULX5 (hereafter P13, ULX1, and ULX5).
First two systems are  two  of the three neutron star ULXs known at present. 
The observations were taken on with {\it XMM-Newton} \citep{jansen2001}
and {\it NuSTAR} \citep{Harrison2013} telescopes simultaneously. 
 Spectral analysis were recently done by \citet{WaltonNGC7793_2017} for P13,
by \citet{israel17} for ULX1, and by \citet{walton13} for ULX5, where their final models contained several components. In this paper we show, that our single model of non-spherical emission from a neutron star with the  multi-color black body emission from an accretion disk fits the broad-band spectra of P13, ULX1 and ULX5 perfectly. 
The normalization of our model constrains  the distance to the source. 
In case of P13, the  resulted distance perfectly
agrees with the distance determination based on the Cepheid method to the hosting galaxy NGC7793 \citep{pietrzynski2010}. Furthermore, the distance to ULX5 from our method
also agrees with the  distance found with radial velocity method to Circinus galaxy
\citep{Koribalski2004}.

\section{Single model of emission and parameters}
\label{sec:mod}

We assume that ULX source is a non-spherical system containing neutron star with the 
accretion disk around it.  The global disk model is not specified here, and the total disk 
emission is parametrized by multi-color black body shape. This approach is correct, since we 
do not know exact accretion efficiency and we deal with objects where standard accretions 
does not account for their high luminosity. For such object geometry, the infinitesimal energy $d \mathcal{F_{\nu}}$ measured by a distant observer is defined as: 
$d \mathcal{F_{\nu}} = I_{\nu} d\omega$,
where $ d\omega $ is a solid angle in steradian [sr] \citep{mihalas1978}. 
This formula is applicable for the flat space, as we assume in this paper. 
In case when the emitter is located close to  black hole, both general and special relativistic corrections should be taken into account as in \citet[][Eq. A4]{fabian1989}. 
Integrating the above formula over the solid angle subtended by the source, 
we obtained energy dependent intensity: $ \mathcal{F_{\nu}}$ as seen by the observer.
This quantity diminishes with the increasing distance and  it is not an intrinsic property of the source. 
 In case of non-spherical systems the observed intensity should be integrated over emitting region individually depending on the source geometry. 
In this paper, we consider emitting region with cylindrical symmetry. 

\begin{figure*}[h!]
 \includegraphics[scale=0.25]{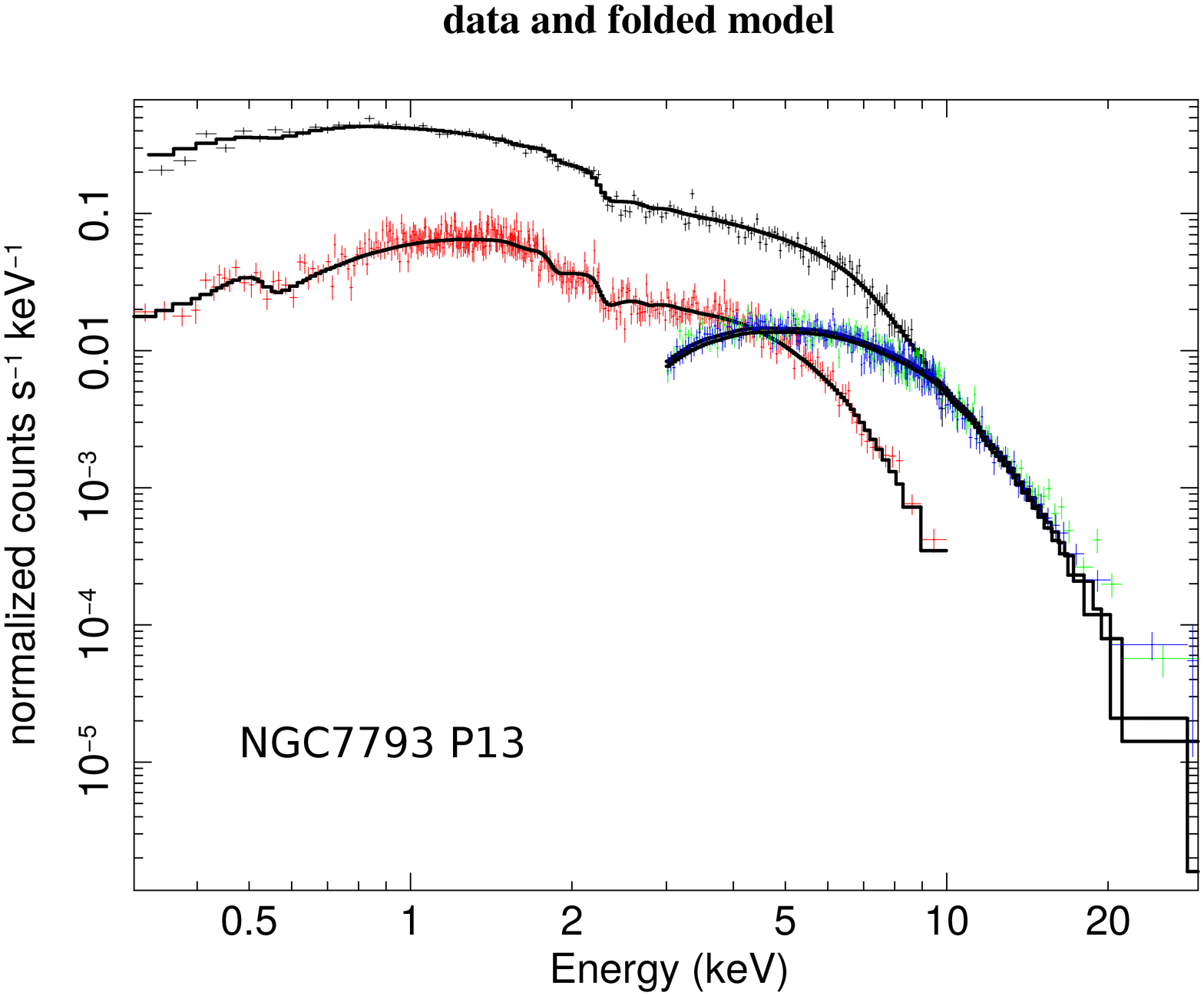}
 \includegraphics[scale=0.25]{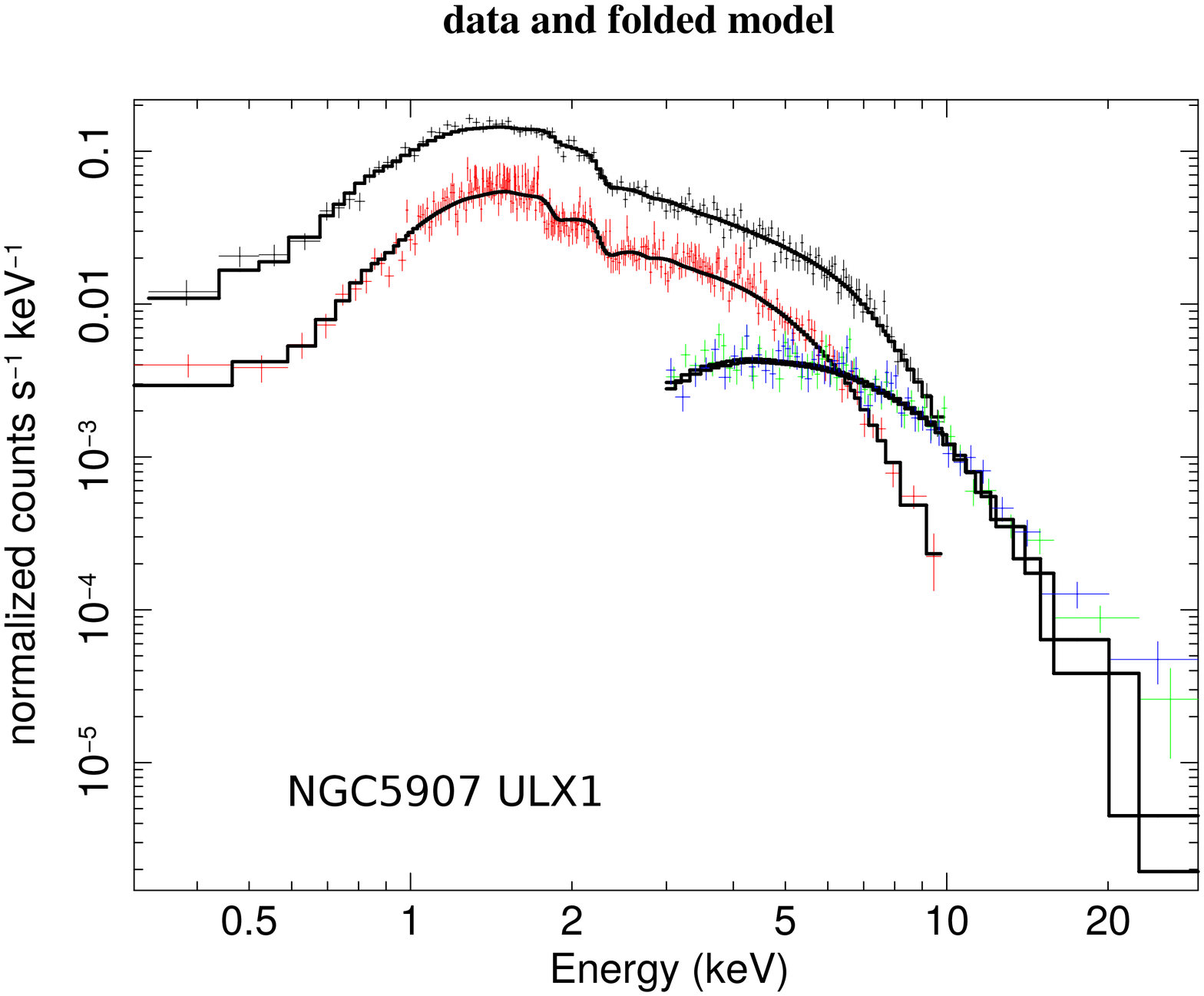}
 \includegraphics[scale=0.25]{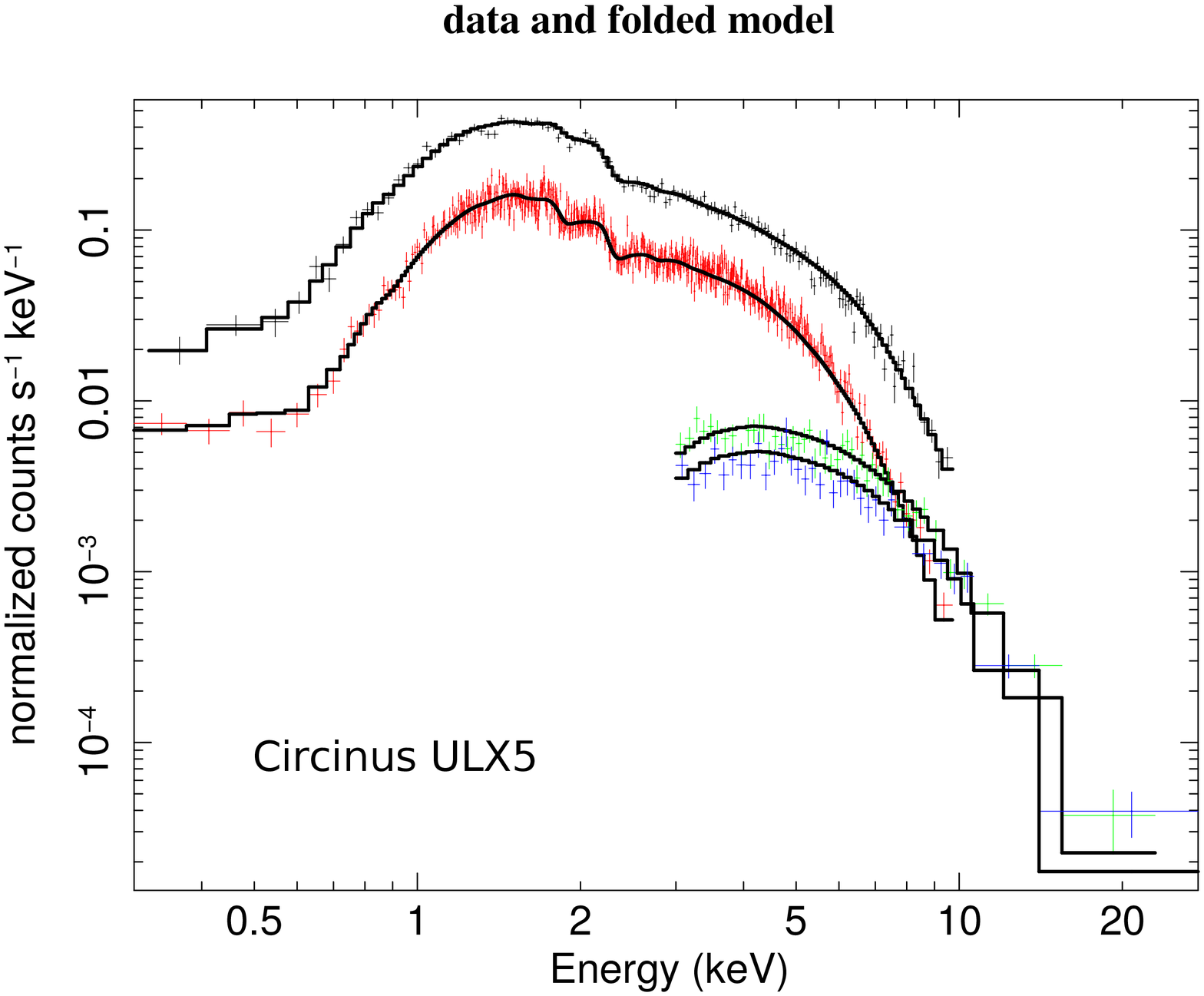}
 \caption{Normalized counts from all detectors used in our spectral fitting analysis
 for P13, ULX1, and ULX5 respectively. 
 Black and red crosses correspond to the {\it XMM-Newton} detectors 
 EPIC-pn and EPIC-MOS. Green and blue crosses are data from
 {\it NuSTAR} FPMA and FPMB respectively. Black solid lines are the best fitted models.}
 \label{fig:counts}
\end{figure*}

In the case of emission from the whole system i.e. neutron star with the accretion disk
around it, we have the contribution from different emitting parts
 \citep[for details see:][]{rozanska17} and 
the final observed energy dependent intensity directed
to the observer is computed analytically as: 
\begin{eqnarray}
\mathcal{F_{\nu, \rm All}} & = &  \left( {1 \over D} \right)^2 \,\, \left[ \pi R_{\rm NS} ^2 \, \left( 
\int_0^1 I_{\nu} \mu d\mu + \int_{\cos{\theta^{\prime}}}^1 I_{\nu} \mu d\mu \right) \right.  \nonumber \\
& +  &   2  \left(  \int_0^{R_{\rm NS}} I_{\nu} \sin{\theta^{\prime}} \, \sqrt{R_{\rm NS}^2-x^2} \, dx  \right.   \nonumber \\ 
& -&  \left. \int_0^{R_{\rm NS} \sin{\theta^{\prime}}} I_{\nu}  \sqrt{R_{\rm NS}^2 \sin^2{\theta^{\prime}}-x^2}\, dx  \right)  \nonumber \\
&+& \pi \sin{\theta^{\prime}}  \left. \left( \int_{R_{\rm in}}^{R_{\rm out}} I_{\nu} R dR 
+  \int_{R_{\rm boost}}^{R_{\rm out}} I_{\nu} R dR \right) \right],
\label{eq:all}
\end{eqnarray}
where $R_{\rm NS}$ is the neutron star radius, $D$ - the distance to the system, 
and $I_{\nu}$ - the  specific intensity emitted from the source surface towards the observer. Other variables denote respectively: $\mu =  \cos{\theta}$, where $\theta$ is 
the angle between direction of the light beam and the normal to the neutron star surface, 
$\theta^{\prime}$ is viewing angle related to the disk inclination as 
$i=90^{\circ}- \theta^{\prime}$
(see Fig.~3 in \citet{rozanska17}), and 
 $x$ is  the variable of integration.
 
First two parts of above equation  correspond to the emission 
from the neutron star surface taking into account the occultation by the disk, 
while second two parts describe the  multi-color black body disk emission with 
eventual occultation by the neutron star. 
Due to  mutual occultation one part of the  disk is integrated 
over radius from the innermost stable orbit $R_{\rm in}$ all the way up to the 
outer radius $R_{\rm out}$, while the second part of the disk - from $R_{\rm boost}= R_{\rm NS} / \sin{\theta^{\prime}}$ up to the outer radius $R_{\rm out}$. $R_{\rm boost}$
the radius up to which neutron star covers the inner disk. 
The above formula was derived analytically  by \citet{rozanska17}, 
where we have demonstrated that the broad-band emitted spectrum from a non-spherical 
system depends on the viewing angle, since  both: multi-color black body disk emission and 
neutron star emission change with angle. 

We assume, that the neutron star radiation  equals  the black body intensity at the given effective temperature $I_{\nu}= B_{\nu} (T_{\rm eff,NS})$. Furthermore, we assume that the emission at different disk radii equals to the local 
Planck function  $I_{\nu}=B_{\nu}(T_{\rm eff}(R))$, with the effective temperature  
given by  a standard multi-color black body  formula:
$ T_{\rm eff}(R)= T_{\rm in} (R/R_{\rm in})^{-p}$,  
where $T_{\rm in}$ is the inner disk temperature and exponent $p$ equals 3/4 for the standard disk. 
Following earlier approach to the spectral fitting of ULXs 
\citep{walton13,bachetti14, furst16,israel17,WaltonNGC7793_2017}, we do 
not connect here the inner disk temperature with an accretion rate, since this 
relation strongly depends on the assumed global disk model and the accretion efficiency. 
Since the latter quantity is not well known and since we expect super Eddington accretion in ULXs, we continue disk parametrization by the inner disk temperature. 
Nevertheless, 
we note here, that our model is useful  for systems where the angle dependent 
specific intensity is given as the results of the radiative transfer calculations 
\citep{madej89,madej1991,hubeny2001,davis2005,rozanska2011}. We plan to 
implement atmospheric models in the future work. 

For the purpose to compare our model to the observed X-ray spectrum of ULX source 
P13, we constructed the grid of models for arbitrarily assumed parameters. 
The non-rotating neutron 
star  has a canonical mass 1.4~M$_{\odot}$,  radius 
12~km, and 11 various effective temperatures, ranging from $2\times 10^6$~K up 
to $4\times 10^7$~K.  The disk local multi-color emission was computed assuming 11 inner disk temperatures  ranging from $2.24 \times 10^{6}$ up to $1.26 \times 10^{7}$~K. 
The grid of $\theta^{\prime}$ angles spans from $10^{\circ}$ up to $90^{\circ}$. 
The lowest value of this angle corresponds  to the almost ``edge on'' disk, whereas the highest value to ``face on'' disk. 

For each disk model, we calculated multi-black body spectrum from 
$R_{\rm in}$ to $R_{\rm out}$ in the range 3-1000~$R_{\rm Schw}$, 
where the $R_{\rm Schw}=2GM_{\rm NS}/c^2$. 
The inner disk radius,  and therefore the inner disk temperature, 
can change due to the: large 
value of magnetic field, strong boundary layer, and when the relativistic corrections are 
taken into account. We plan to include them in the future paper together with full ray tracing procedure. 

Our model normalized by the distance of $D=10$~kpc, was prepared as a table model in the 
FITS (Flexible Image Transport System) format \citep{wells81}, and for the purpose of this paper we named it {\it nsmcbb} (neutron star multi-color black body). 
Final table of total emission from the system
is parametrized by four parameters which will be determined during fitting procedure: 
neutron star effective temperature, $T_{\rm eff,NS}$,  inner disk temperature 
$T_{\rm in}$, viewing angle $\theta^\prime$, and 
normalization $N$. Since the normalization of our model is proportional to $1/D^2$, the 
value of this normalization which comes out from the fitting procedure is 
the direct indicator of the distance to the system, which equals $10/\sqrt{N}$ kpc. 
Below we show that in case of P13 
the distance derived from our fitting is in excellent agreement with 
independent distance measurement. 

\section{P13, ULX1, ULX5 and their X-ray observations}
\label{sec:source}

\begin{figure*}
  \includegraphics[scale=0.24]{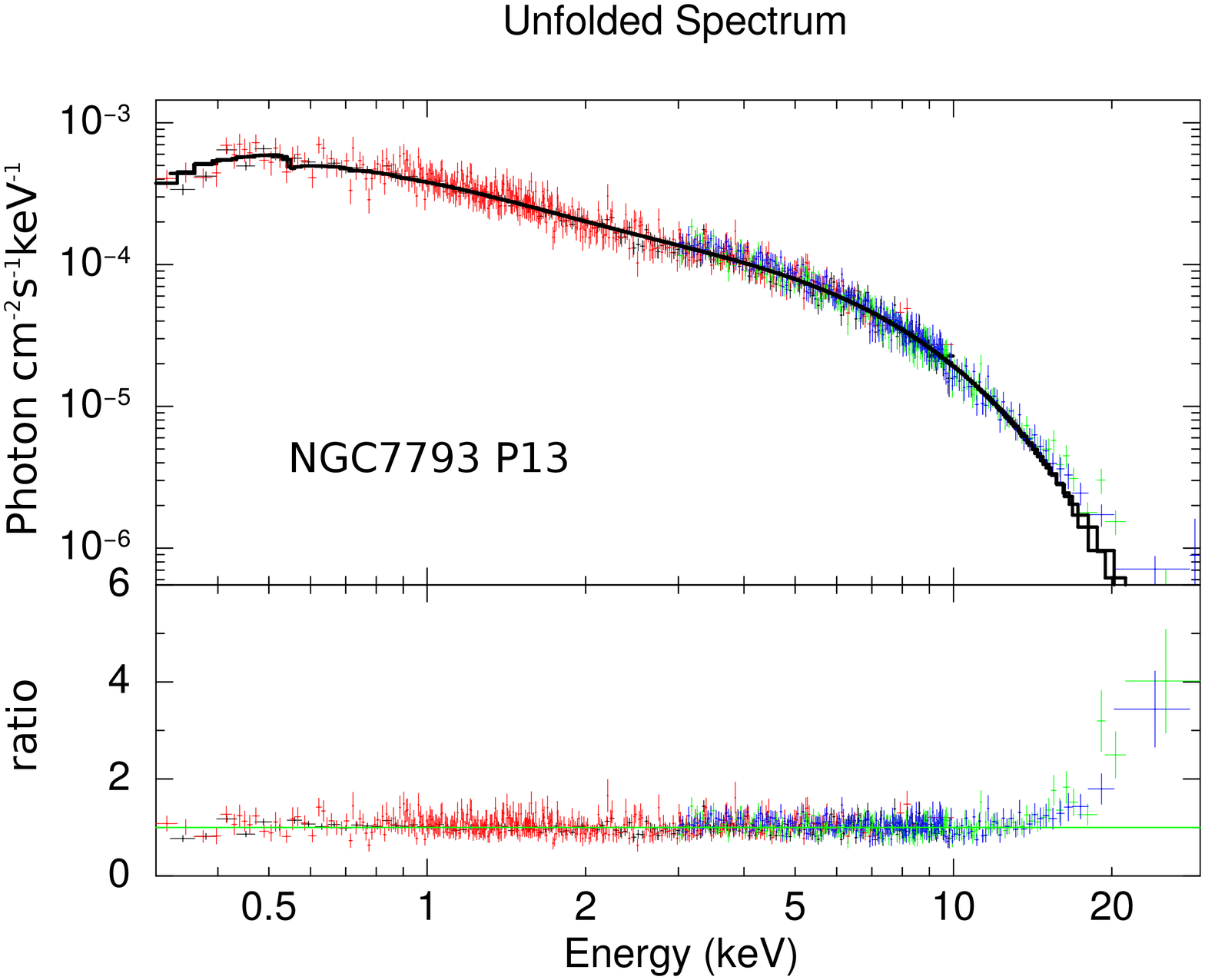}
  \includegraphics[scale=0.24]{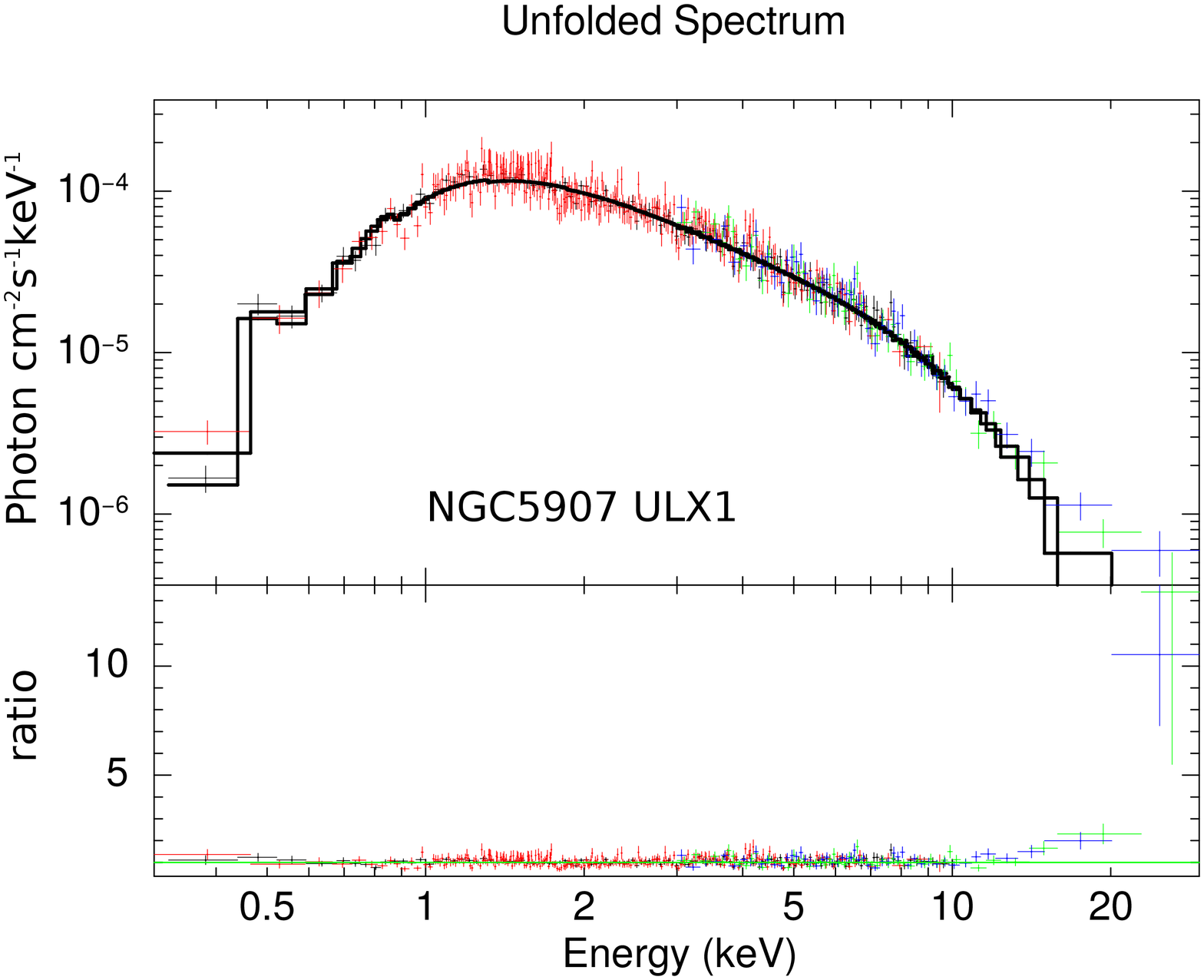}
  \includegraphics[scale=0.24]{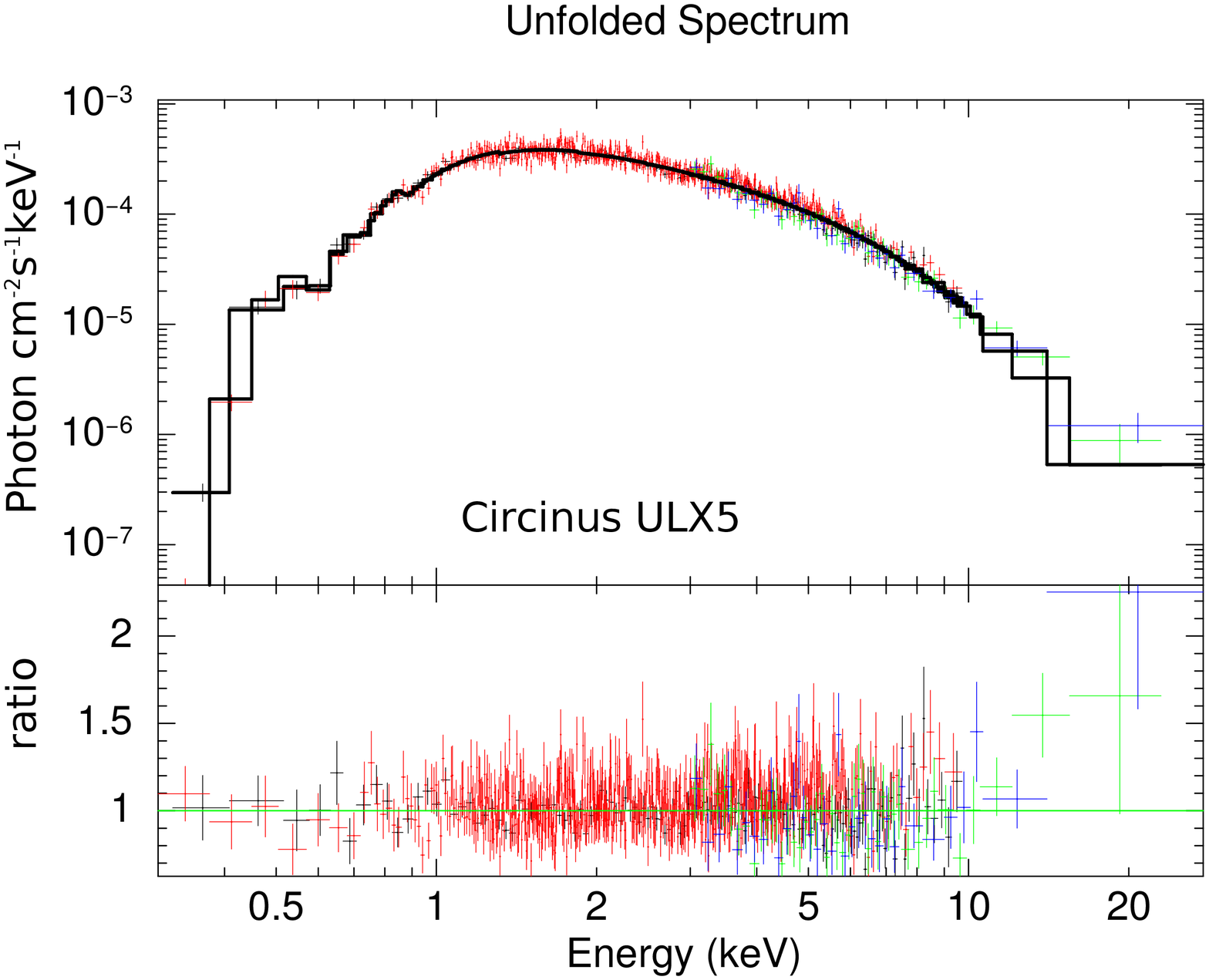}
 \caption{Upper panels shows unfolded photon spectra from all detectors used in our spectral fitting analysis, while lower panels present the ratio i.e. data divided by model. 
All colors have the same meaning as in Fig.~\ref{fig:counts}.
}
 \label{fig:photon}
\end{figure*}

P13  is a variable ULX source, that has been proposed to harbour a stellar
 black hole of a mass less than 15 M$_{\odot}$  \citep{motch2014}.
However, recent studies by \citet{furst16,israel17} have discovered that P13  hosts an accreting  neutron star with the spin period of 0.42 s.  
The source is a part of the  binary system, where it circulates around B91a star of 18-23 
M$_{\odot}$, with the period of 64 days \citep{motch2014}.
Previously, P13 has been reported to reach luminosities from 
$\sim 2 \times 10^{39}$ detected in 1979 by {\it Einstein} satellite up to 
 $\sim 10^{40}$ erg s$^{-1}$ reported recently with join {\it XMM-Newton}
 and {\it NuSTAR} data by \citet{WaltonNGC7793_2017}. The source 
 is rather unobscured with galactic warm absorption value: 
 $N_{\rm H} = 9.60 \pm 0.01 \times 10^{20}$~cm$^{-2}$ \citep{israel17}.
 On the other hand  \citet{WaltonNGC7793_2017} in their paper divided this 
 absorption between galactic absorption with the value of neutral hydrogen 
 estimated towards the object 
 on $N_{\rm H} = 1.2 \times 10^{20}$~cm$^{-2}$ \citep{kalberla2005}, and 
 intrinsic source absorption resulting from the fit : $N_{\rm H} = 8  \pm 1 \times 10^{20}$~cm$^{-2}$. 
P13 is located in NGC7793 galaxy which is a part of Sculptor Group. 
The first distance measurement to the source was done by \citet{karachentsev2003}
as a distance to the galaxy itself and was estimated to be $3.91 \pm 0.41$ Mpc.
Further distance derivation was done within  ARAUCARIA project \citep{gieren2005}
where for the first time Cepheid Variables were detected in the Sculptor Group. 
Cepheids distance to the galaxy NGC7793 was found to be $3.4 \pm 0.17$ Mpc
\citep{pietrzynski2010}.

ULX1 located in NGC5907 galaxy is a part of binary system, with its period being 78 days. 
Recent studies by \citet{israel17} have discovered that ULX1 harbours a 
neutron star at its center, with spin period ranging from 1.13s to 1.43s. 
First estimations of the distance from Earth amounted to 13 Mpc \citep{Tully2009}. 
However, later studies discovered that ULX1 may be located even further form Earth, 
with its distance totaling 17 Mpc \citep{Tully2016}. 
 Such a high distance results in peak luminosities reaching values up 
to $6 \times 10^{40}$ erg s$^{-1}$ as for 13 Mpc \citep{furst17}, which is many times higher then a luminosity for a standard neutron star. The typical values of galactic warm absorption toward 
the ULX1 given by those authors equal from 5.7 to $8.5 \times 10^{21}$~cm$^{-2}$.

ULX5 is located in the outskirts of the Circinus galaxy, with its distance estimated to 
be 2.78 Mpc \citep{Koribalski2004} (distance to the galaxy from radial velocity method).
The first observations of ULX5 at hard X-ray have been made and 
analyzed by \citet{walton13}. Depending on the fitted model, the values of galactic warm absorption toward the source were from 1 to $9.5 \times 10^{21}$~cm$^{-2}$. 
Its peak luminosity was calculated 
 to reach value  $1.6 \times 10^{40}$ erg s$^{-1}$. 
This luminosity was calculated for higher distance of 4~Mpc measured with 
 Tully-Fisher method when Circinus galaxy was discovered \citep{freeman77}. 

In this work we made use of coordinated {\it NuSTAR} (OBSID: 80201010002, 80001042002, 60002039002)  and {\it XMM-Newton} (OBSID: 0781800101, 0729561301, 0701981001)
observations of three sources: P13, ULX1 and ULX5. The observations were taken
on 2016-05-20, 2014-07-09, and 2013-02-03, respectively.  Their exposure times were 
for P13: 118~ks for {\it NuSTAR} and 22/46~ks for EPIC-pn/EPIC-MOS cameras on the board of {\it XMM-Newton} X-ray telescope, for ULX1: 65ks and 38ks/43ks, and for ULX5: 54ks and 24ks/47ks, for the same instruments.

We reduced the {\it NuSTAR} data from both focal plane modules, FPMA and FPMB, by using {\it NuSTAR Data Analysis Software} in accordance to guidelines provided in the {\it NuSTAR Data Analysis Software Guide} (v1.9.2)\footnote{https://heasarc.gsfc.nasa.gov/docs/nustar/analysis/nustar\_swguide.pdf}. Calibration files were taken from the actual database CALDB 
v20170817 throughout the whole process. NUPIPELINE tool was used in order to produce filtered event files, with standard filtering applied. 
We utilized NUPRODUCTS tool to extract source products and instrumental response files form circular regions of radii $60''$ for P13, $40''$ for ULX1, and $55''$ for ULX5. Background products were obtained from four times larger regions located on the same detector chip
as sources.
Besides the basic so called {\it science} data we also included {\it spacecraft science} data following the guideline outlined in {\it NuSTAR Data Analysis Software Guide} \citep[see also][]{walton16}. Therefore we were able to increase our exposure time by approximately 10\% and 
maximazie the signal-to-noise (S/N) ratio. Additional data obtained by this action were merged with standard scientific data by running FTOOLS task ADDASCASPEC. Since we intend to fit 
broad-band data, small errors in responses generated by this tool do not affect our results significantly.
Final spectra, averaged over exposure time, were binned to at  least 20 counts 
per bin for  data fitting process. The extracted normalized counts 
are presented in Fig.~\ref{fig:counts}.

\begin{table}
\begin{center}
\caption{Parameters from fitting  each source with~{\it tbnew*nsmcbb} model. The meaning of  fitted parameters is described in Sec.~\ref{sec:mod}.}
\label{tab:obs}
\begin{tabular}{lllll}
\hline 
Src. & Model & Parameter &  Value & Unit  \\
 \hline \hline
P13 &{\it tbnew}  & $N_{\rm H}$ & $4.77_{-0.44}^{+0.45} \times 10^{20}$ & cm$^{-2}$\\[1.ex]
&{\it nsmcbb} & $T_{\rm eff,NS}$ & $1.819\pm{0.025}$  $\times 10^7$ & K\\ [1.ex]
&{\it nsmcbb} & $T_{\rm in} $  & $1.215_{-0.046}^{+0.036} \times 10^7$ & K\\[1.ex]
&{\it nsmcbb}  & $\theta^\prime$  &  $10\pm{6.59}$ & deg \\ [1.ex]
&{\it nsmcbb}  & $N$  &  $8.62\pm{0.54} \times 10^{-6}$  &  --\\ [1.ex]
\hline
ULX1  &{\it tbnew}  & $N_{\rm H}$ & $4.45_{-0.2}^{+0.13} \times 10^{21}$ & cm$^{-2}$\\[1.ex]
&{\it nsmcbb} & $T_{\rm eff,NS}$ & $1.776^{+0.071}_{-0.032} \times 10^7$ & K\\[1.ex]
&{\it nsmcbb} & $T_{\rm in} $  &  $9.014^{+2.809}_{-0.244} \times 10^6$ & K\\[1.ex]
&{\it nsmcbb}  & $\theta^\prime$  &  $70\pm{1.87}$ & deg \\ [1.ex]
&{\it nsmcbb}  & $N$  &  $2.33_{-0.42}^{+0.70} \times 10^{-6}$  &  --\\ [1.ex]
\hline  
ULX5 &{\it tbnew}  & $N_{\rm H}$ & $5.97_{-0.01}^{+0.01} \times 10^{21}$ & cm$^{-2}$\\[1.ex]
&{\it nsmcbb} & $T_{\rm eff,NS}$ & $1.633^{+0.117}_{-0.109} \times 10^7$ & K\\ [1.ex]
&{\it nsmcbb} & $T_{\rm in} $  &  $1.261^{+0.398}_{-0.075} \times 10^7$ & K\\ [1.ex]
&{\it nsmcbb}  & $\theta^\prime$  &  $12.49^{+1.74}_{-0.71}$ & deg \\ [1.ex]
&{\it nsmcbb}  & $N$  &  $15.23\pm{0.94} \times 10^{-6}$  &  --\\ [1.ex]
\hline 
\end{tabular}
\end{center}
\end{table}

The {\it XMM-Newton} data were reduced with the {\it XMM-Newton Science Analysis
System} (SAS) v16.0.0\footnote{https://www.cosmos.esa.int/web/xmm-newton/sas}, following standard guidelines outlined in the science
analysis threads. We generated the calibrated and concatenated event lists by
running EPCHAIN task for EPIC-pn and EMCHAIN task for both EPIC-MOS modules. Next, we created
an event file and then subjected it to filtering for background flaring with
the help of SAS tasks \textit{evselect}  and \textit{tabgtigen}. To produce final event files 
with spectrum, the source products 
were extracted from circular regions of radii: $36''$ for P13, 
36"/45" for ULX1 (PN/MOS), and 22.5"/27" for ULX5 (PN/MOS),  and background products were extracted from nine times larger regions on the same CCD chip. All regions were extracted in the way to avoid the CCD borders. In \textit{evselect}, we used the filters FLAG$==0$ PATTERN$<4$ for EPIC-pn and FLAG$==0$ PATTERN$<12$ for the EPIC-MOS cameras. The appropriate response files and ancillary files were generated using the SAS commands \textit{rmfgen} and \textit{arfgen} respectively.  Lastly, we combined the spectra form both EPIC-MOS detectors by running FTOOLS task ADDASCASPEC.
Final EPIC-pn spectrum was be binned to at least 10 counts per energy bin, while spectra 
from both MOS cameras were binned to at least 20 counts per energy 
bin for data fitting process. The extracted normalized counts are presented 
in Fig.~\ref{fig:counts}. 

\section{Spectral analysis}
\label{sec:results}

\begin{figure*}
  \includegraphics[scale=0.25]{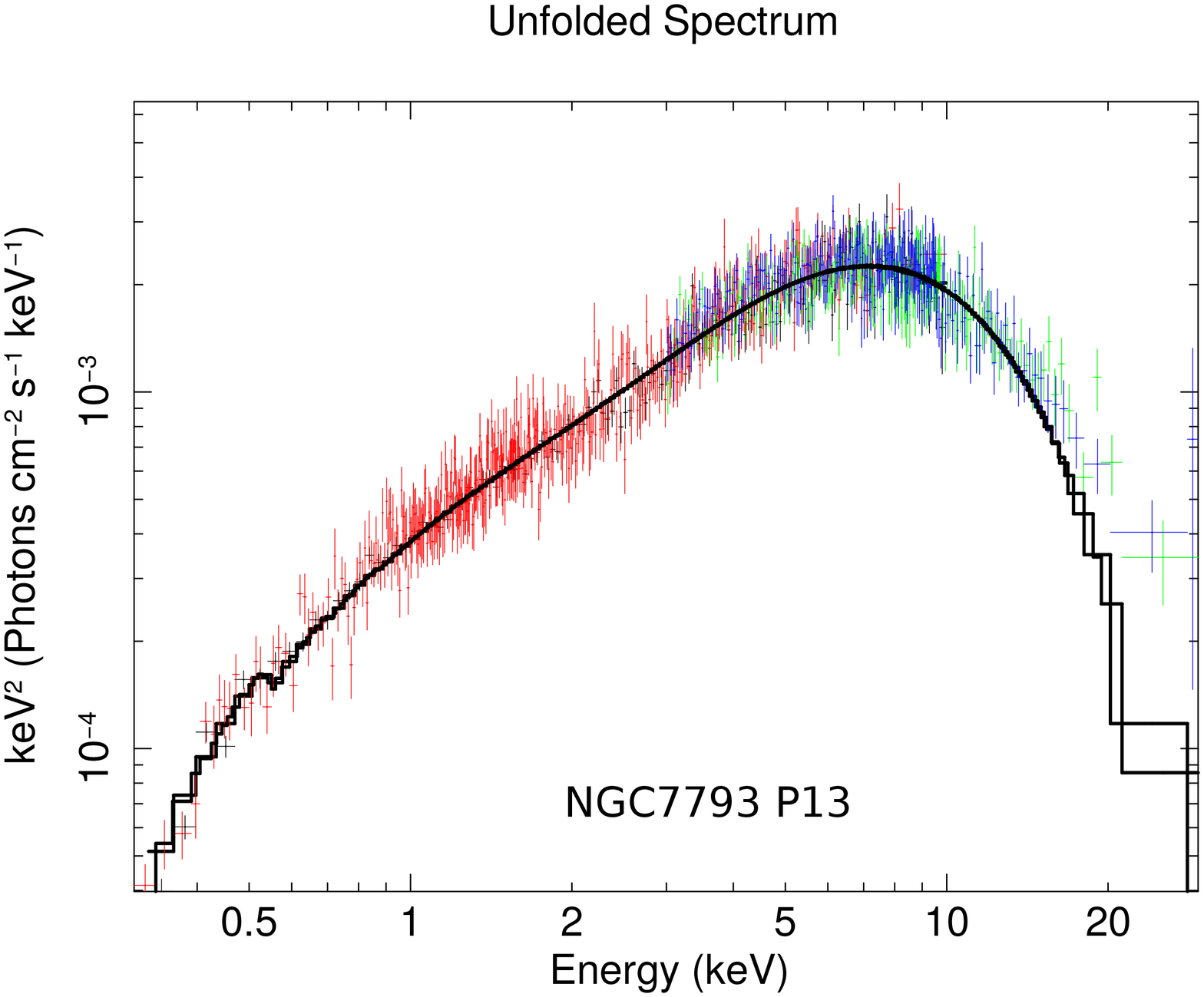}
  \includegraphics[scale=0.25]{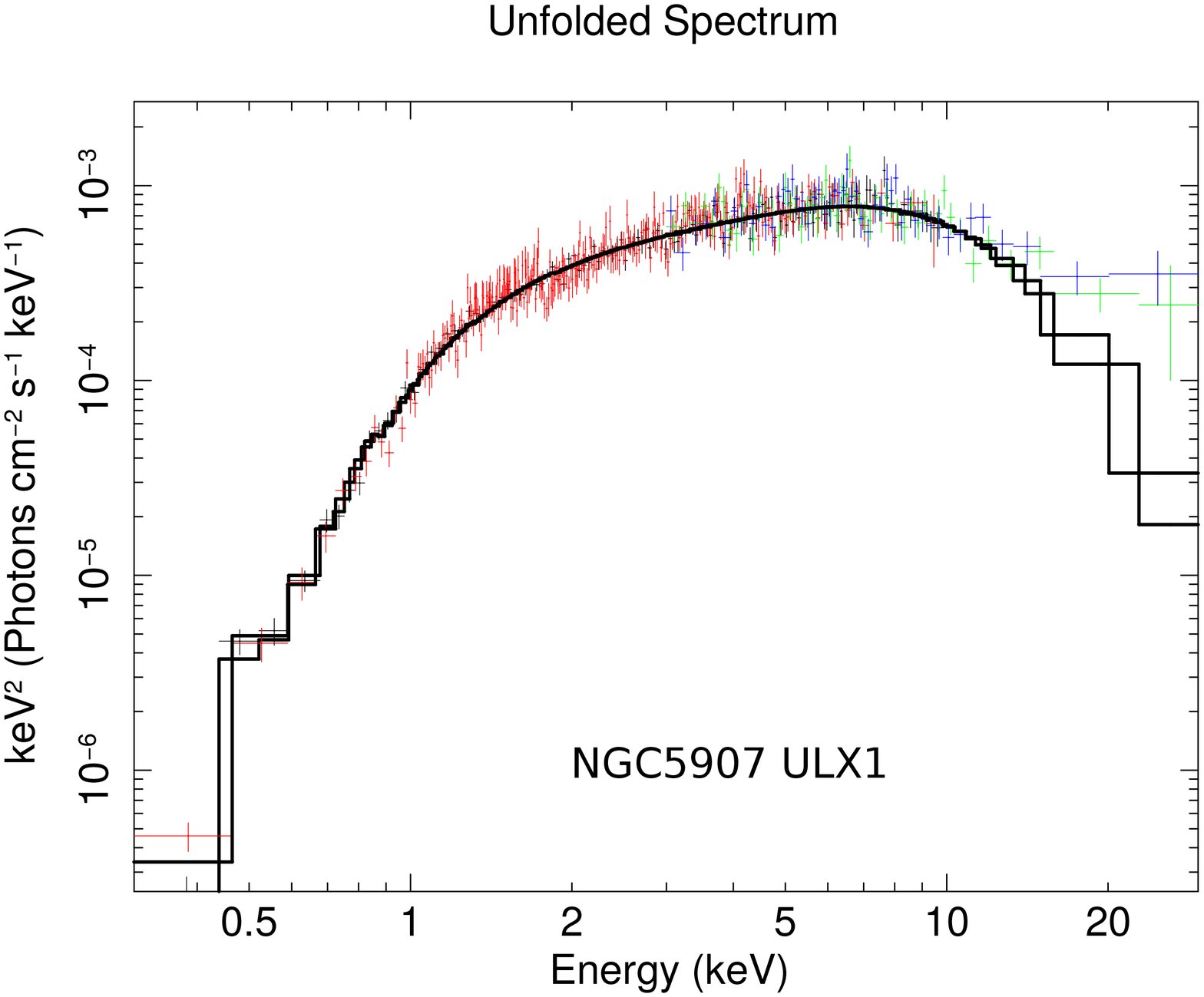}
  \includegraphics[scale=0.25]{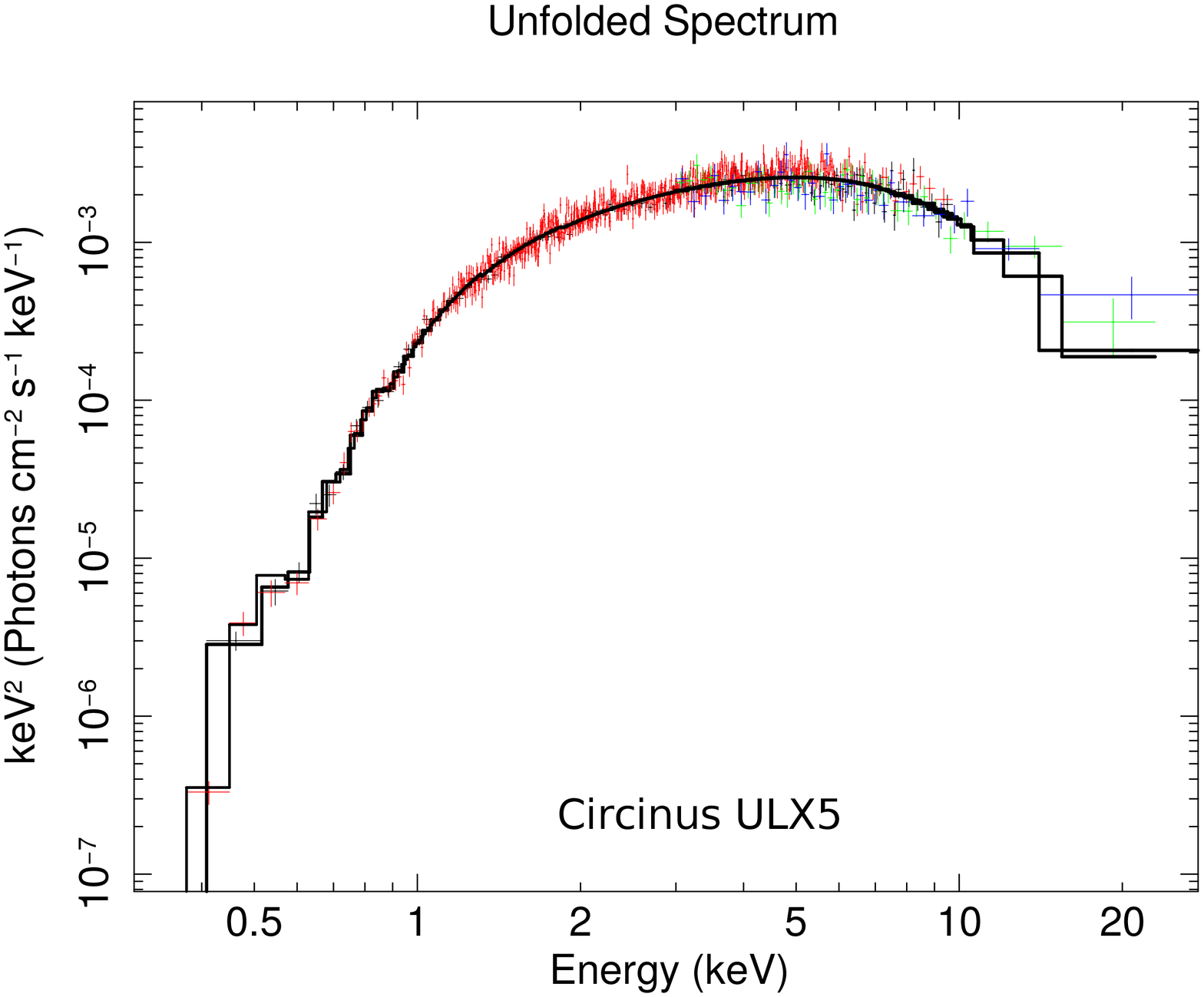}
 \caption{Unfolded  energy spectrum from all detectors used in our spectral fitting analysis. 
 $E*F_{\rm E}$ quantity is plotted to show the maximum emission from 
 hard energy tail which is associated with the emission. 
 Black and red crosses correspond to the {\it XMM-Newton} detectors 
 EPIC-pn and EPIC-MOS. Green and blue crosses are data from
 {\it NuSTAR} FPMA and FPMB respectively. Black solid lines are the best fitted models.}
 \label{fig:spec}
\end{figure*}

We performed spectral fitting of P13, ULX1 and ULX5 data with a single model 
of emission {\it nsmcbb}, presented in Sec~\ref{sec:mod} with the use of 
{\sc xspec} fitting package, version 12.9.0\footnote{https://heasarc.gsfc.nasa.gov/xanadu/xspec/}. In case of each source, all sets of  data were fitted simultaneously with {\it nsmcbb} model multiplied by galactic absorption model {\it tbnew}. All values of metal abundances in 
 {\it tbnew} model were frozen beside hydrogen column density.
 Tab.~\ref{tab:obs} presents all  fitted parameters and uncertainties 
given at the 90\% of confidence level.

The quality of fits is excellent since the data are very good  giving 
1245, 859, and 762 degrees of freedom ($dof$) for each source respectively. The reduced statistics are: $\chi^2/dof = 1.08$ for P13, 1.01 for ULX1 and 1.14 for ULX5, indicating
that our single model of emission from the non-spherical system perfectly 
agrees with data of each source, which we illustrate at Fig.~\ref{fig:photon}. Unfolded spectra 
from all detectors are broad and correctly agree with the models up to 20 keV. 
There is a small deviation for higher energies resulting in the value of ratio (data/model) 
reaching 4, but this feature is always present in ULX sources broad-band spectral
analysis.  Even in the case of multi-component spectral fitting of P13 by  \citet{WaltonNGC7793_2017}, this ratio is between 2 and 3.5. 
This fact may suggest that the hard energy tail is still not well detected by us and 
furthermore not well modeled. 

\begin{table}[h]
\begin{center}
\caption{Results from fitting  each source  with~{\it tbnew*nsmcbb} model. Unabsorbed fluxes 
are calculated using standard {\sc xspec} command. The distance to the source is determined from the normalization of the model. Finally,  the X-ray luminosity is 
computed.}
\label{tab:obs}
\begin{tabular}{llll}
\hline 
Src. & Parameter & Value & Unit \\ 
\hline\hline
P13 & $\chi^2/dof$ & 1344/1245  & -- \\ [1.ex]
& $F_{\rm X}$(2-10 keV) & 4.36 $\times10^{-12}$ & erg s$^{-1}$ cm$^{-2}$\\[1.ex]
& $F_{\rm X}$(0.3-30 keV) & 6.83 $\times10^{-12}$ & erg s$^{-1}$ cm$^{-2}$\\[1.ex]
& $D=10/\sqrt{N}$ & 3.41$^{+0.11}_{-0.10}$ & Mpc\\ [1.ex]
& $L_{\rm X}$(0.3-30 keV)  & 9.59 $\times10^{39}$& erg s$^{-1}$\\ [1.ex]
\hline 
ULX1 & $\chi^2/dof$ & 867/859 & -- \\ [1.ex]
& $F_{\rm X}$(2-10 keV) &  $1.73 \times10^{-12}$ & erg s$^{-1}$ cm$^{-2}$\\[1.ex]
& $F_{\rm X}$(0.3-30 keV) &  $2.81 \times10^{-12}$ & erg s$^{-1}$ cm$^{-2}$\\[1.ex]
& $D=10/\sqrt{N}$ &  $ 6.55_{-0.81}^{+0.69}$ & Mpc\\ [1.ex]
& $L_{\rm X}$(0.3-30 keV)  &  $1.49 \times10^{40}$ & erg s$^{-1}$\\ [1.ex]
\hline
ULX5 & $\chi^2/dof$ & 872/762  & -- \\ [1.ex]
& $F_{\rm X}$(2-10 keV) & $5.78 \times10^{-12}$ & erg s$^{-1}$ cm$^{-2}$\\[1.ex]
& $F_{\rm X}$(0.3-30 keV) & $9.18 \times10^{-12}$ & erg s$^{-1}$ cm$^{-2}$\\[1.ex]
& $D=10/\sqrt{N}$ & $2.60^{+0.05}_{-0.03}$ & Mpc\\ [1.ex]
& $L_{\rm X}$(0.3-30 keV)  & $7.49 \times10^{39}$ & erg s$^{-1}$\\ [1.ex]
\hline
\end{tabular}
\end{center}
\end{table}

The fitted values of galactic absorption are consistent with previous estimations 
given in Sec.~\ref{sec:source},
and since those  values are quite low they do not affect other parameters.
The neutron star effective temperatures are high, giving rise to the hard energy bump 
in the observed spectrum. It is clearly seen in the unfolded energy spectrum 
plotted in $E*F_{\rm E}$ versus $E$ at Fig.~\ref{fig:spec}.
In case of P13 and ULX5, the resulting inclination angles  suggest that the whole systems are observed edge on, and the neutron star is strongly covered by the accretion disk. 
Different case is ULX1, where orientation is almost face on, and therefore occultation is not 
very strong. For all sources considered in this paper, the integration over such emitting area fully 
explains the shape of X-ray observed spectrum.

Since the emission from non-spherical region fits observations, 
we can calculate distances to  all sources from the 
model normalization.  In case of ULX1, the previous distance estimations are uncertain, 
giving two values 13 and 17~Mpc. Distance to ULX1  resulting from  our modeling is a factor of two lower  $D=6.55_{-0.81}^{+0.69}$~Mpc. 

In case of  ULX5, the situation is much better since the distance to the source obtained by our method $D=2.60^{+0.05}_{-0.03}$~Mpc agrees with 2.78~Mpc obtained from radial velocity method by 
\citet{Koribalski2004} (assuming Hubble constant $H_0=75$~km~s$^{-1}$~Mpc).
However, the best agreement we achieved in case of P13, where the obtained value $D=3.41^{+0.11}_{-0.10}$~Mpc from our modeling fully agrees 
with two earlier distance estimations to the 
Sculptor Group being $3.91 \pm 0.41$ Mpc \citep{karachentsev2003}, and 
to the host galaxy NGC7793  from Cepheid method -- $3.4 \pm 0.17$~Mpc
\citep{pietrzynski2010}. It shows that 
our model  is correct and the proper emission from several regions with 
mutual occultation should be taken account
during broad-band data analysis of accreting systems. 

\section{Discussion}
\label{sec:sum}
In this paper we showed,  that the broad-band spectra of  ULX sources: P13 in NGC7793, 
ULX1 in NGC5907, and ULX5 in Circinus galaxy are well fitted by a single model component. 
The fit statistics is excellent: $\chi^2/dof = 1.08$ for P13, 1.01 for ULX1 and 1.14 for ULX5.
The model results from integration over non-spherical and non-uniform emitting 
region.  The best fitted model indicates that all ULXs  have the hot neutron star 
in their centers, with  the disk with high inner temperatures. 
In case of each source  inclination of the whole system was determined.
Our model does not exclude the presence of hot regions, as hot corona, in the 
center since we can observe weak deviation of the model from the data at 20 keV.  Nevertheless, we aimed to show the effect of non-spherical emission in the simplest way, and 
we claim that it fully works for ULXs analyzed in this paper.

Our results clearly show that the integration over true intensity emitted  by a given surface should be applied to explain emission from non-spherical systems. When this is done, we obtain a new possibility to explain broad-band 
spectra by a single model component. Next, we can derive the distance to the source from the model normalization. We derived the distance to P13, $D=3.41^{+0.11}_{-0.10}$~Mpc,
which  is in very good agreement with the Cepheids distance $3.4 \pm 17$~Mpc \citep{pietrzynski2010}. In case of ULX1, the distance  resulting from our modeling is 
$D=6.55_{-0.81}^{+0.69}$~Mpc, a factor of two lower than previous estimates 
\citep{Tully2009,Tully2016}. But both previous estimations were reporting two different numbers of
13 and 17 Mpc distance to this source. 
The distance to  ULX5 obtained by our method 
$D=2.60^{+0.05}_{-0.03}$~Mpc is in agreement with  2.78~Mpc given by 
\citet{Koribalski2004} with radial velocity measurements.

There exists a more general result of our analysis. Any additional soft X-ray bump, which is  very often observed in X-ray spectra of accreting objects  is usually fitted by separate model components, either disk or black-body.
Here, we have proven that this problem can be solved, when we 
integrate emission over the disk with hot inner source. Such central hot source may be for instance a neutron star or 
a hot corona and may be partially attenuated by the disk.

\section*{Acknowledgments}
This research was supported by Polish National Science 
Center grants No. 2015/17/B/ST9/03422, 2015/18/M/ST9/00541, and 2016/21/N/ST9/03311.  This research has made use of data obtained with {\it NuSTAR}, a project
led by Caltech, funded by NASA and managed by NASA/JPL, and
has utilized the NUSTARDAS software package, jointly developed
by the ASDC (Italy) and Caltech (USA). This work has also made
use of data obtained with {\it XMM-Newton} directly funded by ESA Member States.

\bibliographystyle{aa}
\bibliography{refs}

\end{document}